\documentclass{article}
\usepackage{epsfig}

\newcommand{\R}{{I\!\!R}}

\newcommand{\pa}{\partial}
\newcommand{\be}{\begin{equation}}
\newcommand{\ee}{\end{equation}}
\newcommand{\ba}{\begin{eqnarray}}
\newcommand{\ea}{\end{eqnarray}}
\newcommand{\nn}{\nonumber}
\newcommand{\la}{\label} 
\newcommand{\ep}{\Delta t}
\newcommand{\T}{{\cal T}} 
 
\newcommand{\e}{{\rm e}} 
\newcommand{\bp}{{\bf p}}
\newcommand{\bq}{{\bf q}}

\newcommand{\bac}{{\bf a}}

\newcommand{\ct}{{\cal T}}

\newcommand{\bx}{{\bf x}}
\newcommand{\bv}{{\bf v}}

\def\R{{\rm I}\! {\rm R}}

\usepackage{amsmath,amssymb}
\usepackage{amsfonts}
\usepackage{makeidx}

\newtheorem{algorithm}{Algorithm}[section]

\newtheorem{definition}{Definition}[section]

\newtheorem{remark}{Remark}[section]

\begin{document}

\pagestyle{headings}

\title{Splitting methods for Levitron Problems}
\author{J\"urgen Geiser \thanks{University of Greifswald, Institute of Physics, Felix-Hausdorff-Str. 6, D-17489 Greifswald, Germany, E-mail: juergen.geiser@uni-greifswald.de} \and Karl Felix L\"uskow \thanks{University of Greifswald, Institute of Physics, Felix-Hausdorff-Str. 6, D-17489 Greifswald, Germany, E-mail: tt-karl@freenet.de}}

\maketitle

\begin{abstract}
In this paper we describe splitting methods for
solving Levitron, which is motivated to simulate magnetostatic traps 
of neutral atoms or ion traps.
The idea is to levitate a magnetic spinning top 
in the air repelled by a base magnet.
 
The main problem is the stability of the reduced
Hamiltonian, while it is not defined at the
relative equilibrium. Here it is important to derive
stable numerical schemes with high accuracy.
For the numerical studies, we propose novel splitting 
schemes and analyze their behavior.
We deal with a Verlet integrator and improve its accuracy with
iterative and extrapolation ideas. Such a Hamiltonian splitting method, 
can be seen as geometric integrator and saves computational time
while decoupling the full equation system.

Experiments based on the Levitron model are discussed.

\end{abstract}

{\bf Keywords} splitting method, Verlet integrator, iterative and
extrapolation methods, Levitron problem. \\

{\bf AMS subject classifications.} 65M12, 65L06, 65P10.

\section{Introduction}

We are motivated to simulate a Levitron, which is a magnetic spinning top and
can levitate in a magnetic field. 
The main problem of such a nonlinear problem is to achieve a stability
for the calculation of  the critical splint rate.
While the stability of Levitrons are discussed in the work of \cite{dull98}
and their dynamics in \cite{ganz97}, we concentrate on improving the
standard time-integrator schemes for the reduced Hamiltonian systems.
It is important to derive stable numerical schemes with high accuracy
to compute the non-dissipative equation of motions.
For the numerical studies, we propose novel splitting 
schemes and analyze their behavior.
We deal with a standard Verlet integrator and improve its accuracy with
iterative and extrapolation ideas. Such a Hamiltonian splitting method, 
can be seen as geometric integrator and saves computational time
while decoupling the full equation system, see the splitting
ideas in the overview article \cite{chin2011}.

In the following we describe the reduced model of Gans \cite{gans97} and an 
extension based on a novel idea of magentic field of Dullin \cite{dull98} for a 
disk.

\subsection{Hamiltonian of Gans}

In the paper, we deal with the following problem
(reduced Hamiltonian):
\begin{align}
\label{equ1_ham}
\nonumber H=&\frac{1}{2}\left(p_1^2+p_2^2+p_3^2+\frac{p_4^2}{a}+\frac{(p_5-P-6 \cos q_4)^2}{a \sin ^2 q_4} + \frac{p_6}{c} \right) \\
&- M \left[\sin q_4 \left( \cos q_5 \frac{\pa \Psi}{\pa q_1} \sin q_5 \frac{\pa \Psi}{\pa q_2} \right) +\cos q_4 \frac{\pa \Psi}{\pa q_3}  \right] +q_3
\end{align}

The evolution of the dynamical variable $u(\bq,\bp)$ (including $\bq$ and $\bp$ themselves)
is given by the
Poisson bracket,
\begin{equation}
\pa_tu(\bq,\bp)=
                 \Bigl(
		          {{\partial u}\over{\partial \bq}}\cdot
                  {{\partial H}\over{\partial \bp}}
				 -{{\partial u}\over{\partial \bp}}\cdot
                  {{\partial H}\over{\partial \bq}}
				                    \Bigr)=(A+B)u(\bq,\bp).
\label{peq}
\end{equation}

For the non-separable Hamiltonian of (\ref{equ1_ham}), we have:

\begin{eqnarray}
\label{decouple_1}
{\bf \dot q} = \frac{\partial H}{\partial \bp}(\bp,  \bq) = & \left( p_1, p_2, p_3, \frac{p_4}{a}, \frac{\left(p_5 - p_6 \cos q_4 \right)^2}{a \sin^2 q_4}, \frac{p_6 (\cos^2 q_4 + (a /c) \sin^2 _4) - p_5 \cos q4}{a \sin^2 q_4} \right)
\label{ham}
\end{eqnarray}

The same is given for:
 
\begin{align}
\nonumber {\bf \dot p} =& - \frac{\partial H}{\partial \bq}(\bp, \bq) \\
\nonumber &  = ( M\left( \sin q_4 \cos q_5 \frac{\pa^2 \Psi}{\pa q_1^2} + \cos q_4 \frac{\pa^2 \Psi}{\pa q_1 \pa q_3}  \right), M\left( \sin q_4 \cos q_5 \frac{\pa^2 \Psi}{\pa q_2^2} + \cos q_4 \frac{\pa^2 \Psi}{\pa q_2 \pa q_3}  \right), \\
\nonumber &  M\left( \sin q_4 \left(\sin q_5 \frac{\pa^2 \Psi}{\pa q_2 \pa q_3} + \cos q_5 \frac{\pa^2 \Psi}{\pa q_1 \pa q_3} \right) + \cos q_4 \frac{\pa^2 \Psi}{\pa q_3^2} \right)-1, \\
\nonumber &  M\left( \cos q_4 \left( \sin q_5 \frac{\Psi}{q_2} + \cos q_5 \frac{\pa \Psi}{\pa q_1} \right) - \sin q_4 \frac{\pa \Psi}{\pa q_3} \right) - \frac{p_6(p_5-p_6 \cos q_4)}{a \sin q_4}-\frac{\cos q_4(p_5-p_6\cos q_4)^2}{a \sin ^3 q_4}, \\
 &  M\left( \sin q_4 \left( \cos q_5 \frac{\pa \Psi}{\pa q_2} - \sin q_5 \frac{\pa \Psi}{\pa q_1} \right) \right),
0  )
\label{decouple_2}
\end{align}

$A$ and $B$ are Lie operators, or vector fields
\be
A= \frac{\partial H}{\partial \bp} \cdot\frac{\pa}{\pa\bq} \qquad B= -  \frac{\partial H}{\partial \bq} \cdot\frac{\pa}{\pa\bp}
\la{shop} 
\ee

The transfer to the operators are given in the following description.

The exponential operators $\e^{h A}$ and $\e^{h B}$ are then just shift operators,
with ${\cal T}_2(h)$ is a symmetric second order splitting method:
\be
{\cal T}_{2, VV}(h)(\ep)=\e^{(\ep/2) B}\e^{\ep A} \e^{(\ep/2) B}.
\la{str2}
\ee
and corresponds to the velocity form of the Verlet algorithm (VV).

Further the splitting scheme:
\be
{\cal T}_{2, PV}(h)(\ep)=\e^{(\ep/2) A}\e^{\ep B} \e^{(\ep/2) A}.
\la{str2}
\ee
and corresponds to the position-form of the Verlet algorithm (PV). \\

See also the derivation of the Verlet algorithm in Appendix \ref{appendix}.

${\cal T}_{2, VV}(h)(\ep) = S_{AB}(h)$, the symplectic Verlet or leap-frog algorithm is given as:

We start with $(\bq_0, \bp_0)^t = (\bq(t^{n}), \bp(t^{n}))^t $:

\begin{eqnarray}
(\bq_1, \bp_1)^t = \e^{h/2 B} (\bq_0, \bp_0)^t & = & ( I - \frac{1}{2} h \sum_i  \frac{\partial H}{\partial \bq}(\bp_i,  \bq_i)  \frac{\partial}{\partial \bp_i}) (\bq_0, \bp_0)^t ,
\end{eqnarray}

\begin{eqnarray}
(\bq_2, \bv_2)^t = \e^{h A} (\bq_1, \bv_1)^t & = &  ( I + h \sum_i  \frac{\partial H}{\partial \bp}(\bp_i,  \bq_i)  \frac{\partial}{\partial \bq_i}) (\bq_1, \bp_1)^t , 
\end{eqnarray}

\begin{eqnarray}
(\bq_3, \bv_3)^t =  \e^{h/2 B} (\bq_2, \bp_2)^t & = & ( I - \frac{1}{2} h \sum_i  \frac{\partial H}{\partial \bq}(\bp_i,  \bq_i)  \frac{\partial}{\partial \bp_i}) (\bq_2, \bp_2)^t  .
\end{eqnarray}

And the substitution is given the algorithm for one time-step $n \rightarrow n+1$ and we obtain:

$(\bq(t^{n+1}), \bv(t^{n+1}))^t = (\bq_3, \bv_3)^t$.

\subsection{Higher order Expansion of Verlet-algorithm}

In the following we extend the Verlet algorithm with
respect to higher order terms.

Such terms are important in the application with iterative schemes
to achieve higher order schemes.

We start with $(\bq_0, \bp_0)^t = (\bq(t^{n}), \bp(t^{n}))^t $:

\begin{eqnarray}
(\bq_1, \bp_1)^t & = &\e^{h/2 B} (\bq_0, \bp_0)^t \nonumber  \\
               & = & ( I  + \sum_{j=1}^N \frac{1}{j !}\big( - \frac{1}{2} h \sum_i  \frac{\partial H}{\partial \bq}(\bp_i,  \bq_i)  \frac{\partial}{\partial \bp_i} \big)^j ) (\bq_0, \bp_0)^t , 
\end{eqnarray}

\begin{eqnarray}
(\bq_2, \bv_2)^t & = & \e^{h A} (\bq_1, \bv_1)^t \nonumber \\
               & = & ( I  + \sum_{j=1}^N \frac{1}{j !}\big( h \sum_i  \frac{\partial H}{\partial \bp}(\bp_i,  \bq_i)  \frac{\partial}{\partial \bq_i} \big)^j ) (\bq_0, \bp_0)^t ,
\end{eqnarray}

\begin{eqnarray}
(\bq_3, \bv_3)^t & = & \e^{h/2 B} (\bq_2, \bp_2)^t \nonumber \\
& = & ( I  + \sum_{j=1}^N \frac{1}{j !}\big( - \frac{1}{2} h \sum_i  \frac{\partial H}{\partial \bq}(\bp_i,  \bq_i)  \frac{\partial}{\partial \bp_i} \big)^j ) (\bq_0, \bp_0)^t .
\end{eqnarray}

And the substitution is given the algorithm for one time-step $n \rightarrow n+1$ and we obtain:

$(\bq(t^{n+1}), \bv(t^{n+1}))^t = (\bq_3, \bv_3)^t$.

\section{Iterative Schemes for coupled problems}

Based on the nonlinear equations, we have to deal with linearization
or nonlinear averaging techniques.
In the following, we discuss the fixed point iteration and Newton's method.

We solve the nonlinear problem:
\begin{eqnarray}
\label{nonlinear}
 F(x) = 0 ,
\end{eqnarray}
where $F: \R^n \rightarrow \R^n$.

\subsection{Fixed-point iteration}

The nonlinear equations can be formulated as fixed-point problems:
\begin{eqnarray}
\label{fix}
 x = K(x) ,
\end{eqnarray}
where $K$ is the fixed-point map and is nonlinear, e.g. $K(x) = x - F(x)$.

A solution of (\ref{fix}) is called fix-point of the map $K$.

The fix-point iteration is given as:
\begin{eqnarray}
\label{fix}
 x_{i+1} = K(x_i) ,
\end{eqnarray}
and is called {\it nonlinear Richardson iteration}, {\it Picard iteration},
or {\it the method of successive substitution}.

\begin{definition}
Let $\Omega \le \R^n$ and let $G: \Omega \rightarrow \R^m$.
$G$ is Lipschitz continuous on $\Omega$ with Lipschitz constant $\gamma$ if 
\begin{eqnarray}
\label{fix_2}
|| G(x) - G(y) || \le \gamma || x - y || ,
\end{eqnarray}
for all $x, y \in \Omega$.
\end{definition}

For the convergence we have to assume that $K$ be a contraction map on $\Omega$
with Lipschitz constant $\gamma < 1$.

\begin{algorithm}

We apply the fix-point iterative scheme to decouple the
non-separable Hamiltonian problem (\ref{decouple_1}) and (\ref{decouple_2}).

\begin{eqnarray}
& &{\bf \dot q}_i = \frac{\partial H}{\partial \bp}(\bp_{i-1},  \bq_{i-1}) ,  t \in [t^n, t^{n+1}] \\
& & {\bf \dot p}_i = - \frac{\partial H}{\partial \bq}(\bp_{i-1}, \bq_{i-1}) , t \in [t^n, t^{n+1}] \\
&& p(t^0) = p_0, q(t^0)= q_0 , 
\end{eqnarray}
the starting solutions for the i-th iterative steps
are given as : \\
$\bp_{i-1}(t), \bq_{i-1}(t)$ are the solutions of the $i-1$ th iterative step\\
and we have the initial condition for the fix-point iteration: \\
$ (\bp_0(t),  \bq_0(t))^t = (\bp(t^n),  \bq(t^n))^t$

We assume that we have convergent results after $i = 1 \ldots, m$ iterative steps or with the stopping criterion:

$ \max (|| \bp_{i+1} - \bp_{i} || , || \bq_{i+1} - \bq_i ||) \le err $,

while $|| \cdot ||$ is the Euclidean norm (or a simple vector-norm, e.g. $L_2$). 

\end{algorithm}

Iterative Verlet applied to the Hamiltonian (\ref{decouple_1}) and (\ref{decouple_2}) :

We start with $(\bq_0, \bp_0)^t = (\bq(t^{n}), \bp(t^{n}))^t $:

The iterative scheme is given as:

\begin{eqnarray}
\bq_{i}(t) & = & \bq(t^n) + h \frac{\partial H}{\partial \bp }(\bp(t^n) - \frac{h}{2} \frac{\partial H}{\partial \bq}(\bp_{i-1}(t), \bq_{i-1}(t)), \bq(t^n)) , \\
\bp_{i}(t) & = & \bp(t^n) - \frac{h}{2} \frac{\partial H}{\partial \bq}(\bp_{i-1}(t), \bq_{i-1}(t))  \nonumber \\
&& - \frac{h}{2} \frac{\partial H}{\partial \bq}\Bigg( \bigg( \bp(t^n) - \frac{h}{2} \frac{\partial H}{\partial \bq}(\bp_{i-1}(t), \bq_{i-1}(t)) \bigg), \nonumber \\
&&  \bq(t^n) + h \frac{\partial H}{\partial \bp }\bigg( \bp(t^n) - \frac{h}{2} \frac{\partial H}{\partial \bq}(\bp_{i-1}(t), \bq_{i-1}(t)), \bq(t^n) \bigg) \Bigg) ,\\
&& \mbox{for} \; t \in [t^n, t^{n+1}], h= t^{n+1} - t^n, n=0,1,\ldots, N, \\
&& \; i = 1,2, 3\ldots, I ,
\end{eqnarray}
where $I = 3$ or $4$. 

\vspace{0.5cm}

For the fix-point iteration, we have the problem of the
initialization, means the start of the iterative scheme.
We can improve the starting solution with a preprocessing
method, which derives a first improved initial solution. 

{\bf Improve Initialization Process}

\begin{algorithm}

To improve the initial solution we can start with:

1.) We initialize with a result of the explicit Euler-method :

$(\bq_0, \bp_0)^t = (\bq(t^{n+1})_{Euler 1st}, \bp(t^{n+1})_{Euler 1st})^t $:

2.)  We initialize with a result of the explicit RK-method :

$(\bq_0, \bp_0)^t = (\bq(t^{n+1})_{RK 4th}, \bp(t^{n+1})_{RK 4th})^t $:

\end{algorithm}

\subsection{Newton's method}

We solve the nonlinear operator equation (\ref{nonlinear}).

While $F: D \subset X \rightarrow Y$ with the Banach spaces $X, Y$
is given with the norms $|| \cdot ||_X$ and $|| \cdot ||_Y$.
Let $F$ be at least once continuous differentiable,
further we assume $x_0$ is a starting solution of the
unknown solution $x^*$.  

Then the {\it successive linearization} lead to the
general Newton's method:
\begin{eqnarray}
\label{newton}
 F'(x_i) \Delta x_i = - F(x_i) ,
\end{eqnarray}
where $\Delta x_i = x_{i+1} - x_i$ and $i= 0,1,2, \ldots .$

The method derive the solution of a nonlinear problem by solving
the following algorithm.

\begin{algorithm}

\smallskip By considering the sequential splitting method we obtain
the following algorithm. We apply the equations
\begin{eqnarray}
& &{\bf \dot q} - \frac{\partial H}{\partial \bp}(\bp,  \bq) = 0 ,  t \in [t^n, t^{n+1}] \\
& & {\bf \dot p} + \frac{\partial H}{\partial \bq}(\bp, \bq) = 0 , t \in [t^n, t^{n+1}] \\
&& {\bf p}_n = {\bf p}(t^n), {\bf q}_n= {\bf q}(t^n) , 
\end{eqnarray}
where ${\bf p} = (p_1, \ldots, p_6)^t$ and ${\bf q} = (q_1, \ldots, q_6)^t$ 
into the Newtons-formula we have:
\begin{eqnarray}
F({\bf p}, {\bf q}) = {\bf \dot x} + \frac{\partial H}{\partial \bx}(\bx)
\end{eqnarray}
and we can compute 
\begin{eqnarray}
{\bf x}^{(k+1)} = {\bf x}^{(k)} - D(F({\bf x}^{(k)}))^{-1} F({\bf x}^{(k)}), 
\end{eqnarray}
where $D(F({\bf x}))$ is the Jacobian matrix and $k = 0, 1, \ldots $.

We stop the iterations when we obtain : $|{\bf x}^{(k+1)} - {\bf x}^{(k)}| \le err$,
where $err$ is an error bound, e.g. $err = 10^{-4}$.

The solution vector $F$
is given as:
\begin{eqnarray}
&& F({\bf x}) =
\left(
\begin{array}{c}
F_{q,1}({\bf x}) \\
F_{q,2}({\bf x}) \\
\vdots \\
F_{q,6}({\bf x}) \\
F_{p,1}({\bf x}) \\
F_{p,2}({\bf x}) \\
\vdots \\
F_{p,6}({\bf x}) 
\end{array}
\right)
\end{eqnarray}
where ${\bf x} = (q_1, \ldots, q_6, p_1, \ldots, p_6)^t$ and
\begin{eqnarray}
&& F({\bf x}) =  \left(
\begin{array}{c}
F_{q}({\bf x}) \\
F_{p}({\bf x})
\end{array}
\right) = \left(
\begin{array}{c}
{\bf \dot q} - \frac{\partial H}{\partial {\bf p}} \\
{\bf \dot p} - \frac{\partial H}{\partial {\bf q}}
\end{array}
\right).
\end{eqnarray}

The Jacobian matrix for the equation system is given as :
\begin{eqnarray}
&& D F({\bf x }) = \left(
\begin{array}{c c c c}
\frac{\partial F_1}{\partial x_1} & \frac{\partial F_1}{x_2}& \ldots & \frac{\partial F_1}{\partial x_{12}} \\
\\
\frac{\partial F_2}{\partial x_1} & \frac{\partial F_2}{x_2}& \ldots & \frac{\partial F_2}{\partial x_{12}} \\
\\
\vdots \\
\\
\frac{\partial F_{12}}{\partial x_1} & \frac{\partial F_{12}}{x_2}& \ldots & \frac{\partial F_{12}}{\partial x_{12}} \\
\end{array}
\right)
\end{eqnarray}
where ${\bf x} = (x_1, \ldots, x_{12})^t = (q_1, \ldots, q_6, p_1, \ldots, p_6)^t$.

\end{algorithm}

\section{Splitting Methods}
\label{oper}

In the following, we discuss the different splitting schemes.

The simplest such symmetric product is  
\be
{\cal T}_2(h)=S_{AB}(h) \quad{\rm or}\quad {\cal T}_2(h)=S_{BA}(h).
\ee
If one naively assumes that
\be
{\cal T}_2(h)=\e^{\ep(A+B)}+ Ch^3+Dh^4+\cdots,
\ee
then a Richardson extrapolation would only give
\be
\frac1{k^2-1}\left[k^2 \ct_2^k(h/k)-\ct_2(h)\right]=\e^{\ep(A+B)}+ O(h^4),
\ee
a third-order algorithm. 

 Thus 
for a given set of $n$ whole numbers $\{k_i\}$ one can have a $2n$th-order approximation  
\be
{\rm e}^{\ep(A+B)}
=\sum_{i=1}^n c_i{\cal T}_2^{k_i}\left(\frac\ep{k_i}\right)
+O(h^{2n+1}).
\la{mulexp}
\ee

For orders four to ten, one has explicitly:
\be
{\cal T}_4(\ep)=-\frac13{\cal T}_2(\ep)
+\frac43{\cal T}_2^2\left(\frac\ep{2}\right)
\la{four} ,
\ee
\be
{\cal T}_6(\ep)=\frac1{24} {\cal T}_2(\ep)
-\frac{16}{15}{\cal T}_2^2\left(\frac\ep{2}\right)
+\frac{81}{40}{\cal T}_2^3\left(\frac\ep{3}\right)
\la{six} ,
\ee
\be
{\cal T}_8(\ep)=-\frac1{360} {\cal T}_2(\ep)
+\frac{16}{45}{\cal T}_2^2\left(\frac\ep{2}\right)
-\frac{729}{280}{\cal T}_2^3\left(\frac\ep{3}\right)
+\frac{1024}{315}{\cal T}_2^4\left(\frac\ep{4}\right)
\la{eight} ,
\ee
\ba
&&{\cal T}_{10}(\ep)=\frac1{8640} {\cal T}_2(\ep)
-\frac{64}{945}{\cal T}_2^2\left(\frac\ep{2}\right)
+\frac{6561}{4480}{\cal T}_2^3\left(\frac\ep{3}\right)\nn\\
&&\qquad\qquad\quad-\frac{16384}{2835}{\cal T}_2^4\left(\frac\ep{4}\right)
+\frac{390625}{72576}{\cal T}_2^5\left(\frac\ep{5}\right).
\la{ten}
\ea

\section{Iterative MPE method}

Based on the nonlinear problem, we extend the MPE method to an iterative scheme.

\begin{algorithm}

Iterative Verlet applied to the Hamiltonian (\ref{decouple_1}) and (\ref{decouple_2}) :

We start with $(\bq_0, \bp_0)^t = (\bq(t^{n}), \bp(t^{n}))^t $: \\

The iterative scheme for computing $\T_2(p_i, q_i, h) = ( \bq_{i}(h), \bp_{i}(h)^t  )$ is given as:

\begin{eqnarray}
\T_2(p_i, q_i, h) & = &\e^{h/2 B}  \e^{h A} \e^{h/2 B} (\bq_0, \bp_0)^t \nonumber  \\
               & = & ( I  + \sum_{j=1}^N \frac{1}{j !}\big( - \frac{1}{2} h \frac{\partial H}{\partial \bq}(\bp_{i-1},  \bq_{i-1})  \frac{\partial}{\partial \bp(t^n)} \big)^j ) \\
 && ( I  + \sum_{j=1}^N \frac{1}{j !}\big( h  \frac{\partial H}{\partial \bp}(\bp_{i-1},  \bq_{i-1})  \frac{\partial}{\partial \bq(t^n)} \big)^j ) \nonumber \\
&& ( I  + \sum_{j=1}^N \frac{1}{j !}\big( - \frac{1}{2} h \frac{\partial H}{\partial \bq}(\bp_{i-1},  \bq_{i-1})  \frac{\partial}{\partial \bp(t^n)} \big)^j )  (\bq(t^n), \bp(t^n))^t \nonumber 
\end{eqnarray}

We have the higher order schemes given as:

\be
{\cal T}_4(p_i, q_i, h) = -\frac13{\cal T}_2(p_i, q_i, h)
+\frac43{\cal T}_2^2\left(p_i, q_i, \frac{h}{2}\right)
\la{four}
\ee
\be
{\cal T}_6(p_i, q_i, h)=\frac1{24} {\cal T}_2(p_i, q_i, h)
-\frac{16}{15}{\cal T}_2^2\left(p_i, q_i, \frac{h}{2}\right)
+\frac{81}{40}{\cal T}_2^3\left(p_i, q_i, \frac{h}{3}\right)
\la{six}
\ee
\ba
&& {\cal T}_8(p_i, q_i, h)=-\frac1{360} {\cal T}_2(p_i, q_i, h)
+\frac{16}{45}{\cal T}_2^2\left(p_i, q_i, \frac{h}{2}\right)
\nn\\
&&\qquad\qquad\quad -\frac{729}{280}{\cal T}_2^3\left(p_i, q_i, \frac{h}{3}\right)
+\frac{1024}{315}{\cal T}_2^4\left(p_i, q_i, \frac{h}{4}\right)
\la{eight}
\ea
\ba
&&{\cal T}_{10}(p_i, q_i, h)=\frac1{8640} {\cal T}_2(p_i, q_i, h)
-\frac{64}{945}{\cal T}_2^2\left(p_i, q_i, \frac{h}{2}\right)
+\frac{6561}{4480}{\cal T}_2^3\left(p_i, q_i, \frac{h}{3}\right)\nn\\
&&\qquad\qquad\quad-\frac{16384}{2835}{\cal T}_2^4\left(p_i, q_i, \frac{h}{4}\right)
+\frac{390625}{72576}{\cal T}_2^5\left(p_i, q_i, \frac{h}{5}\right).
\la{ten}
\ea
where $\mbox{for} \; t \in [t^n, t^{n+1}], h= t^{n+1} - t^n, n=0,1,\ldots, N,$ \\
$ i = 1,2, 3\ldots, I ,$

and we have a stopping criterion or a fixed number of iterative steps,
e.g. $I = 3$ or $4$.

\end{algorithm}

\section{Numerical Examples}
\label{appl}

The Levitron is described on the base of rigid body theory. With the convention of Goldstein \cite{goldstein81} for the Euler angles the angular velocity $\omega_{\phi}$ is along the $z$-axis of the system, $\omega_{\theta}$ along the line of nodes and $\omega_{\psi}$ along the $z'$-axis. Transforming them into body coordinates one gets

\be
 \omega =\begin{pmatrix}
 \dot \phi \sin \theta \sin \psi + \dot \theta \cos \psi \\ 
 \dot \phi \sin \theta \cos \psi + \dot \theta \sin \psi \\ 
 \dot \phi \cos \theta + \dot \psi
\end{pmatrix} 
\ee

Finally the kinetic energy can be written as 

\be
T=\frac{1}{2}\left[m(\dot x^2+\dot y^2+\dot z^2) + A( \dot \theta ^2+\dot \phi ^2 \sin ^2 \theta) + C(\dot \psi +\dot \phi \cos \theta)^2 \right]
\ee

The potential energy $U$ is given by the sum of the gravitational energy and the interaction potential of the Levitron in the magnetic field of the base plate:

\be 
U=mgz-\mu (\sin \psi \sin \theta \frac{\Phi}{x} + \cos \psi \sin \theta \frac{\Phi}{y} + \cos \theta \frac{\Phi}{z})
\ee  

with $mu$ as the magnetic moment of the top and $\Phi$ the magneto-static potential. Following Gans \cite{gans97} we uses the potential of a ring dipole as approximation for a magnetized plane with a centered unmagnetized hole. Furthermore we introduced a nondimensionalization for the variables and the magneto-static potential:

\be
\Psi=\frac{Z}{(1+Z^2)^{3/2}}-(X^2+Y^2)\frac{3}{4}\frac{(2Z^2-3)Z}{(1+Z^2)^{7/2}}
\ee

Lengths were scaled by the radius R of the base plane, mass were measured in units of $m$ and energy in units of $mgh$. Therefore the one time unit is $\sqrt{R/g}$.
So the dimensionless Hamiltonian with $\bf q=(X,Y,Z,\theta , \psi , \phi )$ is given by
\begin{align} 
H=&\frac{1}{2}\left(p_1^2+p_2^2+p_3^2+\frac{p_4^2}{a}+\frac{(p_5-P-6 \cos q_4)^2}{a \sin ^2 q_4} + \frac{p_6}{c} \right) \\
&- M \left[\sin q_4 \left( \cos q_5 \frac{\pa \Psi}{\pa q_1} \sin q_5 \frac{\pa \Psi}{\pa q_2} \right) +\cos q_4 \frac{\pa \Psi}{\pa q_3}  \right] +q_3
\end{align}
with $a$ and $c$ as the nondimensionalized inertial parameter and $M$ as the ratio of gravitational and magnetic energy.

\begin{figure}[htbp]
\centering	
	\begin{minipage}[b]{0.47\textwidth}
	\centering 
	\includegraphics[width=\textwidth]{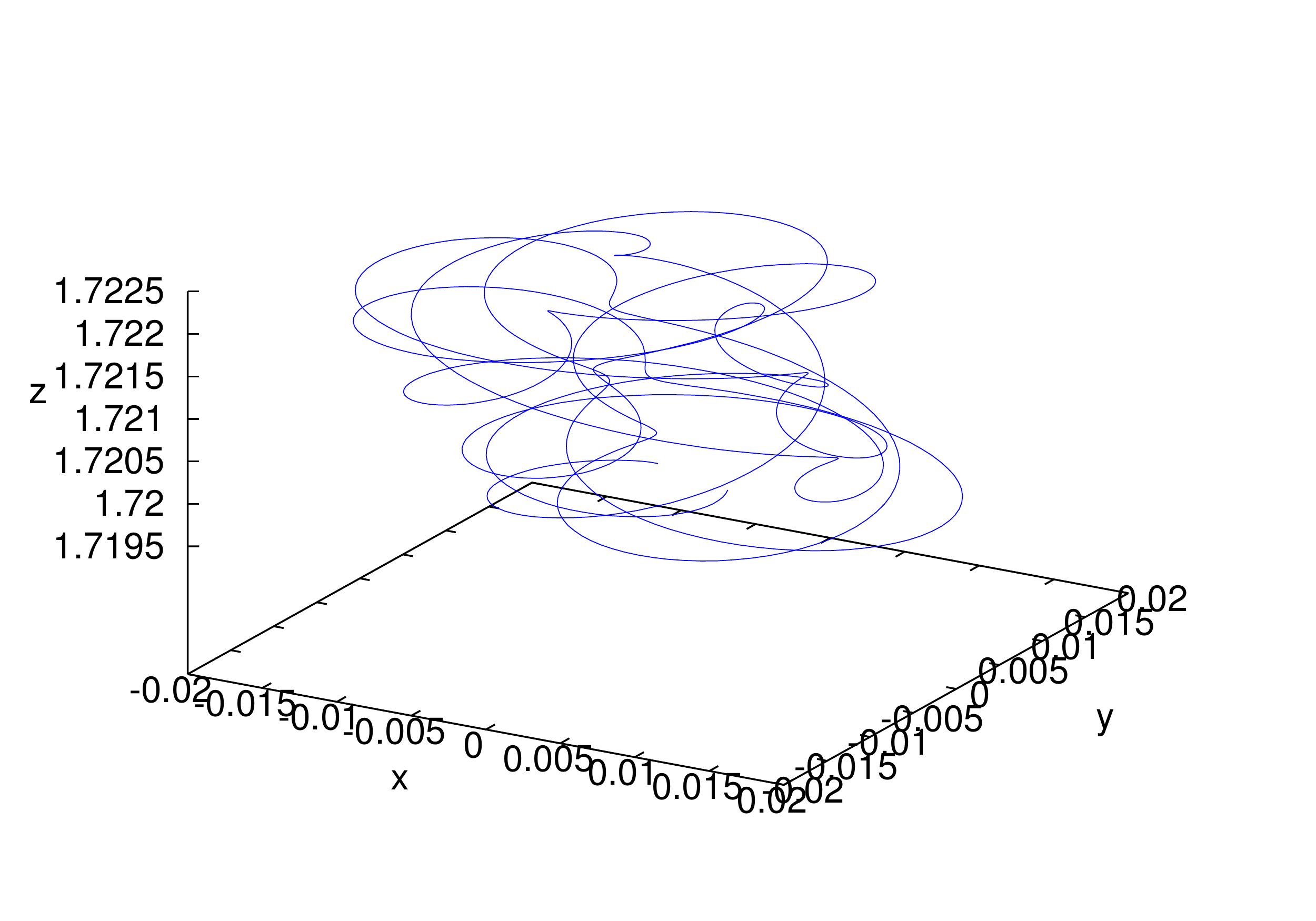}
	\end{minipage}
	\hfill
	\begin{minipage}[b]{0.47\textwidth}
	\centering
	\includegraphics[width=\textwidth]{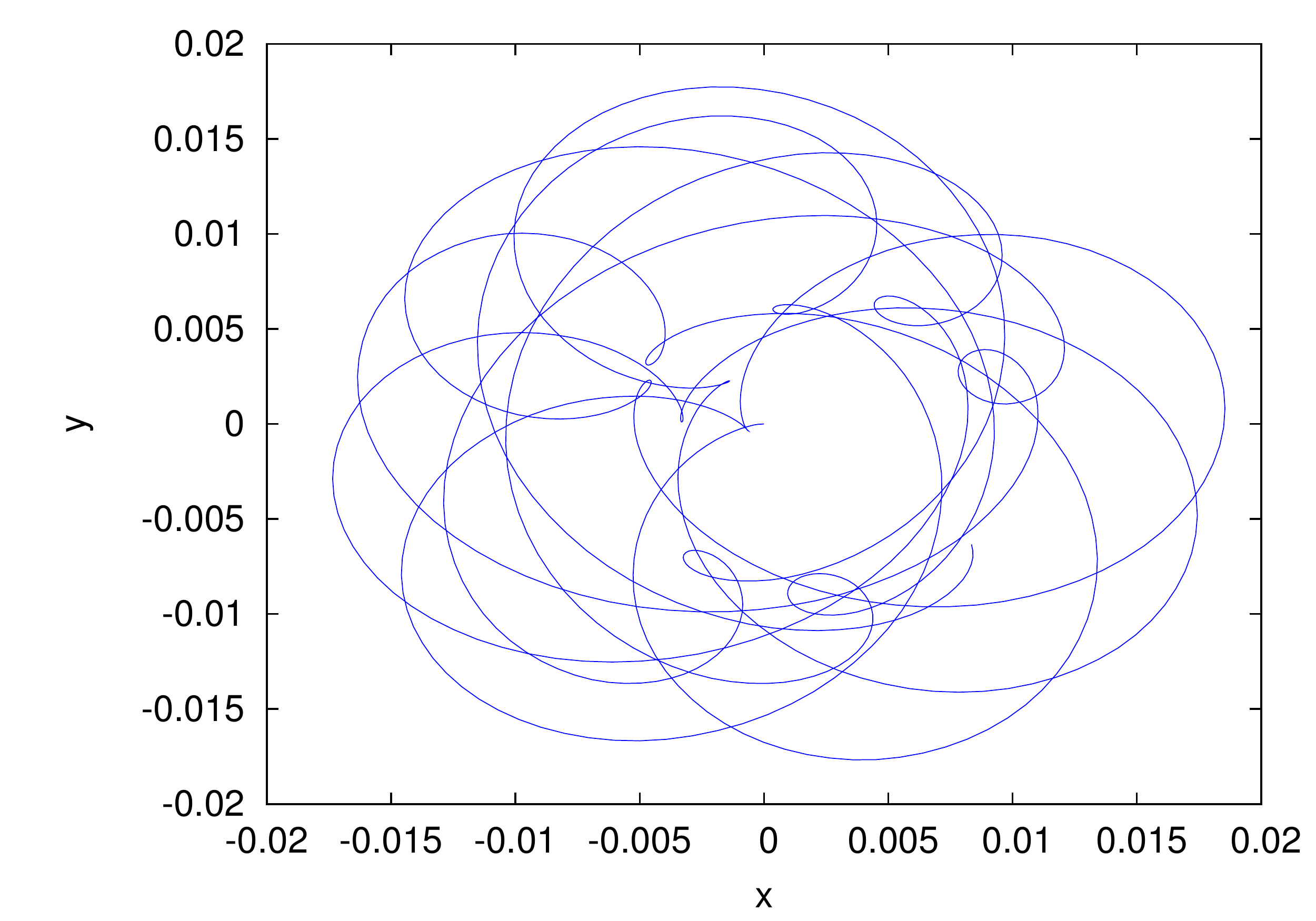}
	\end{minipage}
 \caption{stable trajectory of the center of mass in 3D and 2D} 
 \label{fig:trajectory}
\end{figure}

Solving the equations of motion \eqref{decouple_1} and \eqref{decouple_2} numerically with the methods described above, one is able to plot the movement of the center of mass like it is shown in figure \ref{fig:trajectory}. In the plot the first 5 seconds of a stable trajectory were plotted. With a longer calculation we tested, that the top would levitate for more than one minute. 

The axis in the plot show the nondimensional variables X,Y and Z. The trajectory starts at the equilibrium point $(q_1,q_2,q_3)=(0,0,1.72)$. This trajectory was calculated with the fourth-order Runge–Kutta method with a small timestep of $10^{-5}$ units of time, where one time unit is about $59,5\rm{ms}$, because of the nondimensionalization. For further considerations this trajectory will be used as a reference solution.

\begin{figure}[htbp]
\centering	
	\begin{minipage}[b]{0.47\textwidth}
	\centering 
	\includegraphics[width=\textwidth]{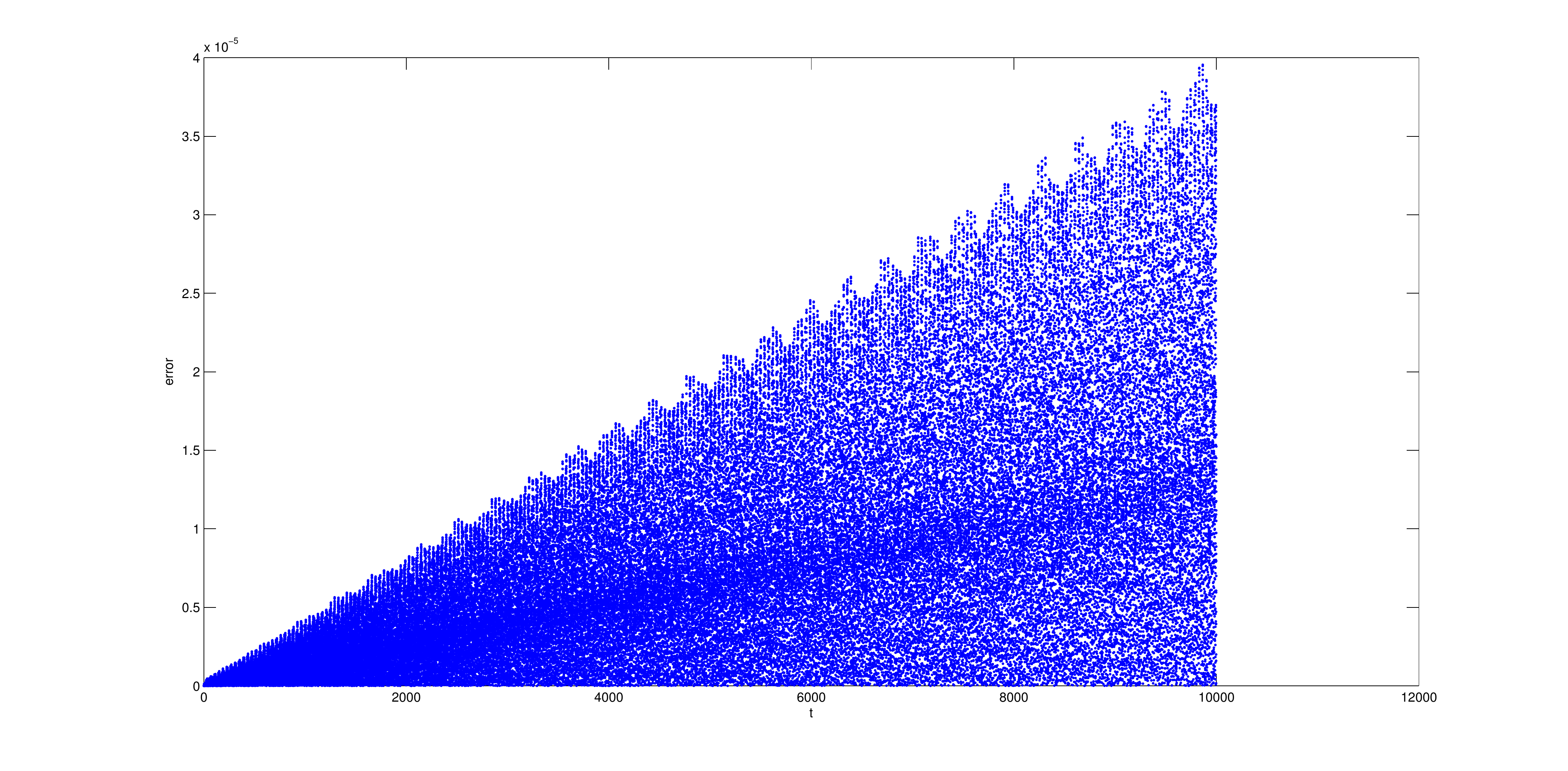}
	\end{minipage}
	\hfill
	\begin{minipage}[b]{0.47\textwidth}
	\centering
	\includegraphics[width=\textwidth]{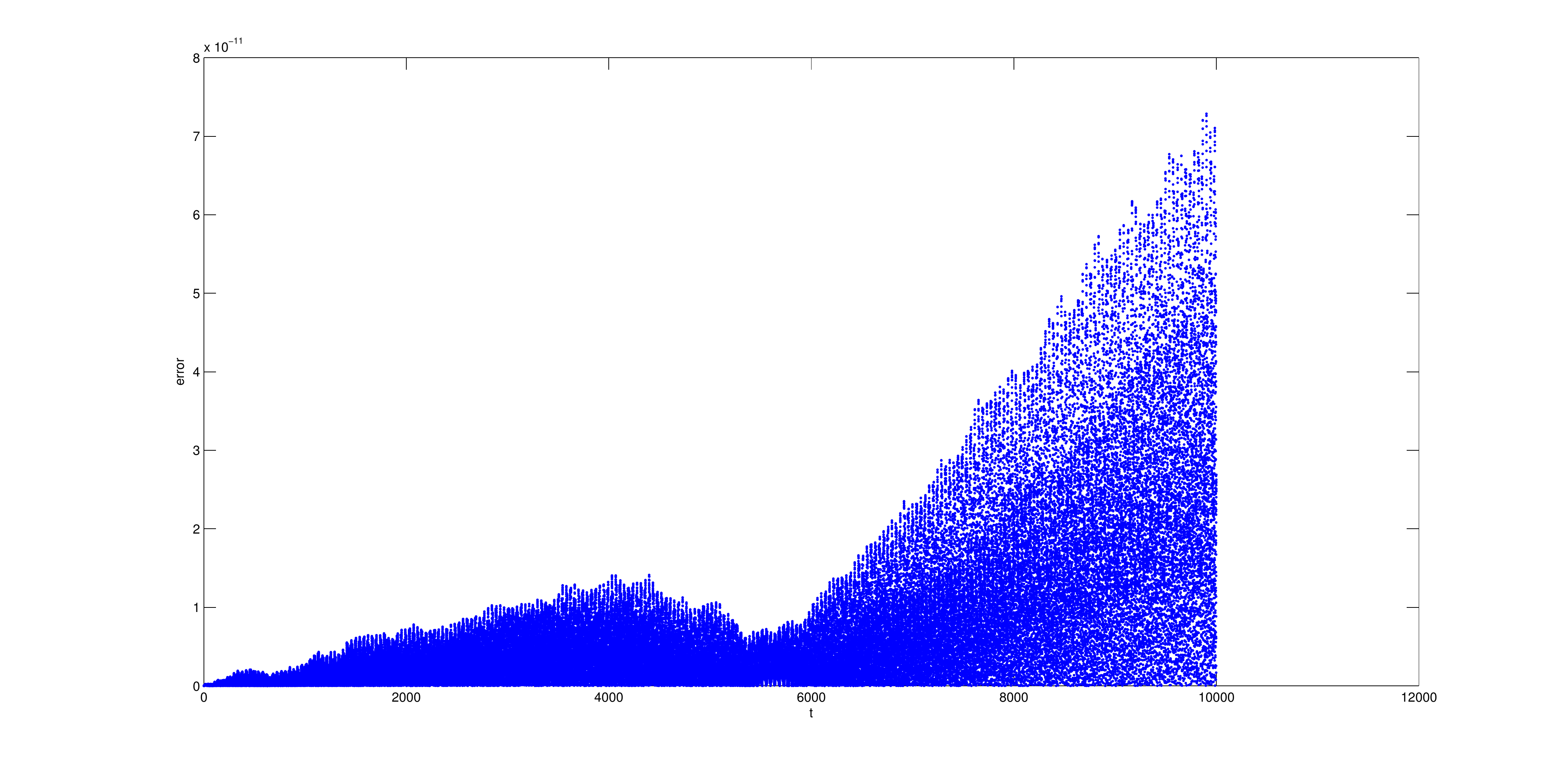}
	\end{minipage}
 \caption{Errors of the numerical scheme: Runge-Kutta method (explicit 4th order).} 
 \label{fig:rk_error}
\end{figure}

The errors of the different time-steps with Runge-Kutta are given in Figure \ref{fig:rk_error}. 

\begin{figure}[htbp]
\centering	
	\begin{minipage}[b]{0.47\textwidth}
	\centering 
	\includegraphics[width=\textwidth,angle=-90]{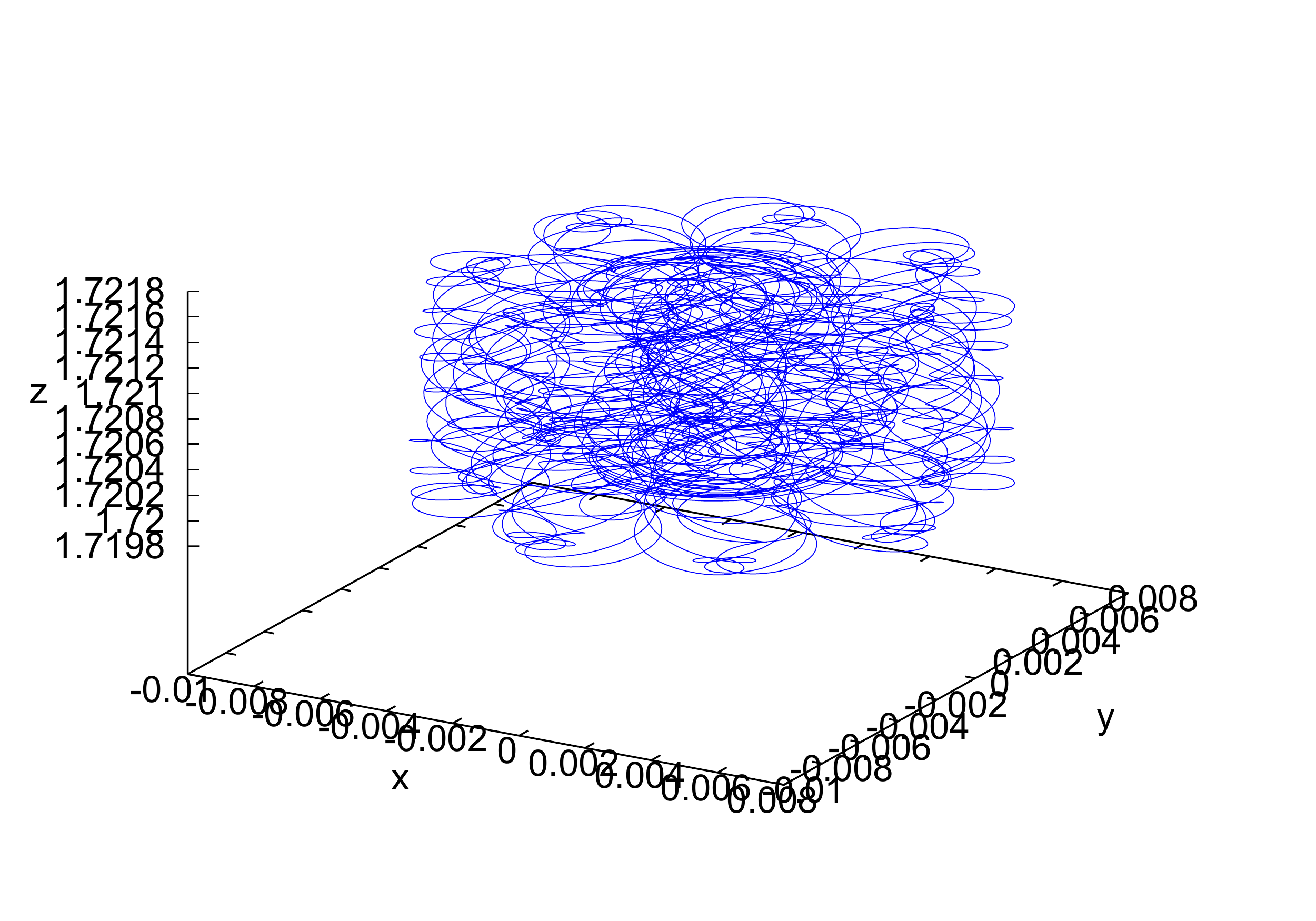}

	\end{minipage}
	\hfill
	\begin{minipage}[b]{0.47\textwidth}
	\centering
	\includegraphics[width=\textwidth,angle=-90]{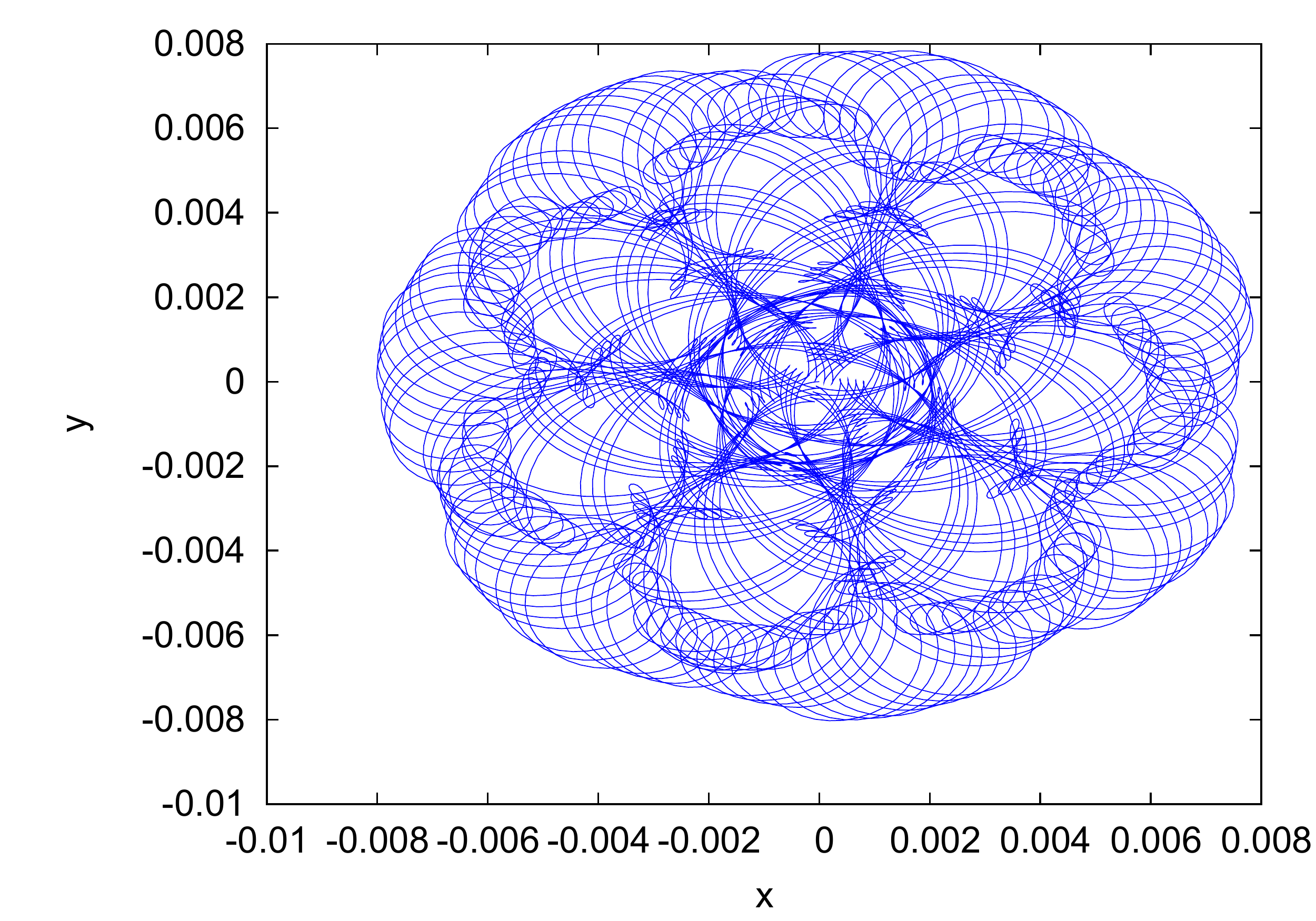}
	\end{minipage}
 \caption{trajectory calculated with Verlet algorithm} 
 \label{fig:trajectory_verlet}
\end{figure}

The same equations were solved with the iterative Verlet algorithm described before. Due to the long computation time needed, we simulated only 1000 timesteps and compare the trajectory with the reference solution from the Runge-Kutta algorithm. In figure \ref{fig:trajectory_verlet} is shown how the trajectory of the same initial conditions looks like with the Verlet algorithm. 

\begin{figure}[htbp]
\centering	
	\begin{minipage}[b]{0.47\textwidth}
	\centering 
	\includegraphics[width=\textwidth]{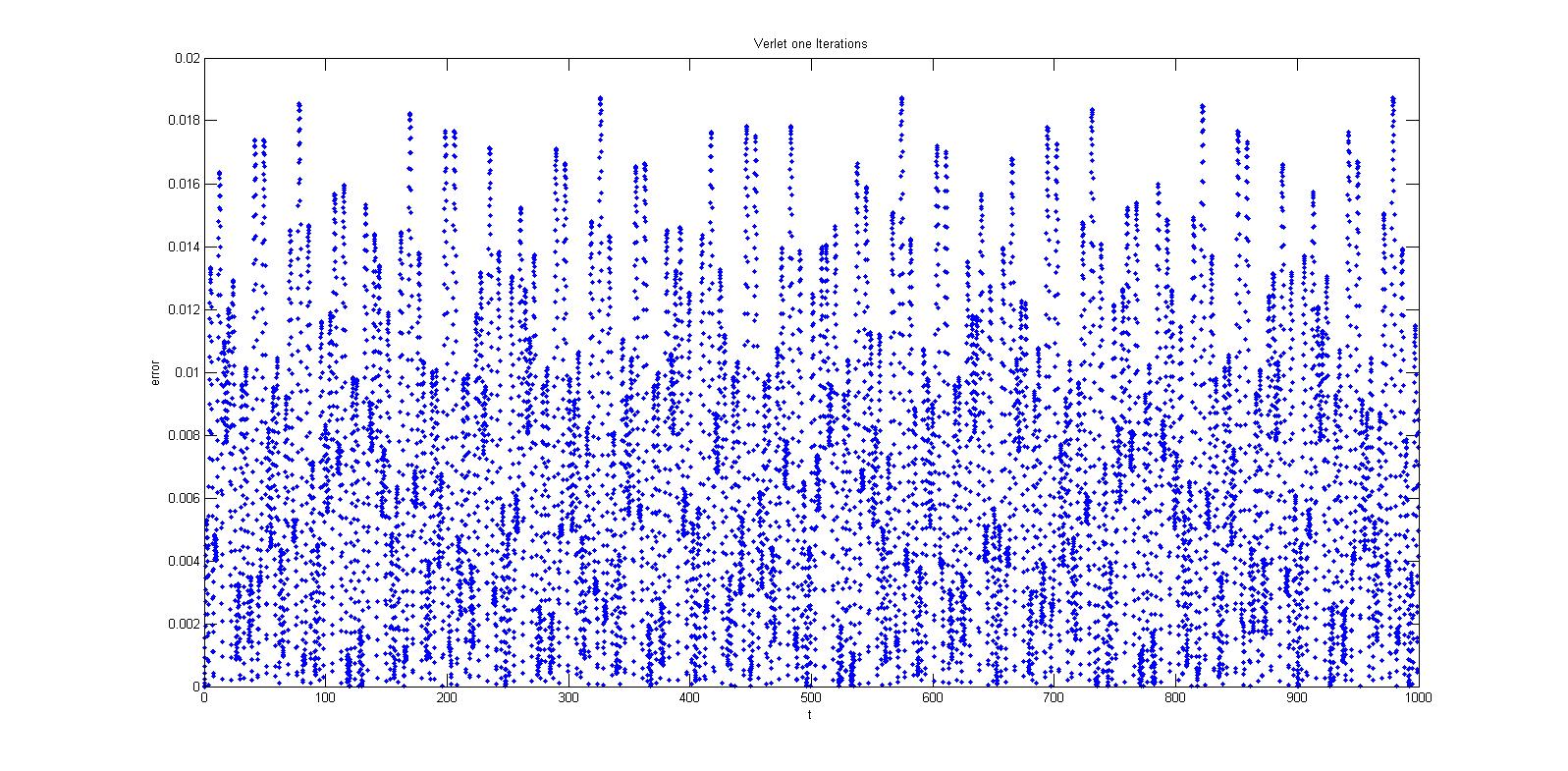}
	\end{minipage}
	\hfill
	\begin{minipage}[b]{0.47\textwidth}
	\centering
	\includegraphics[width=\textwidth]{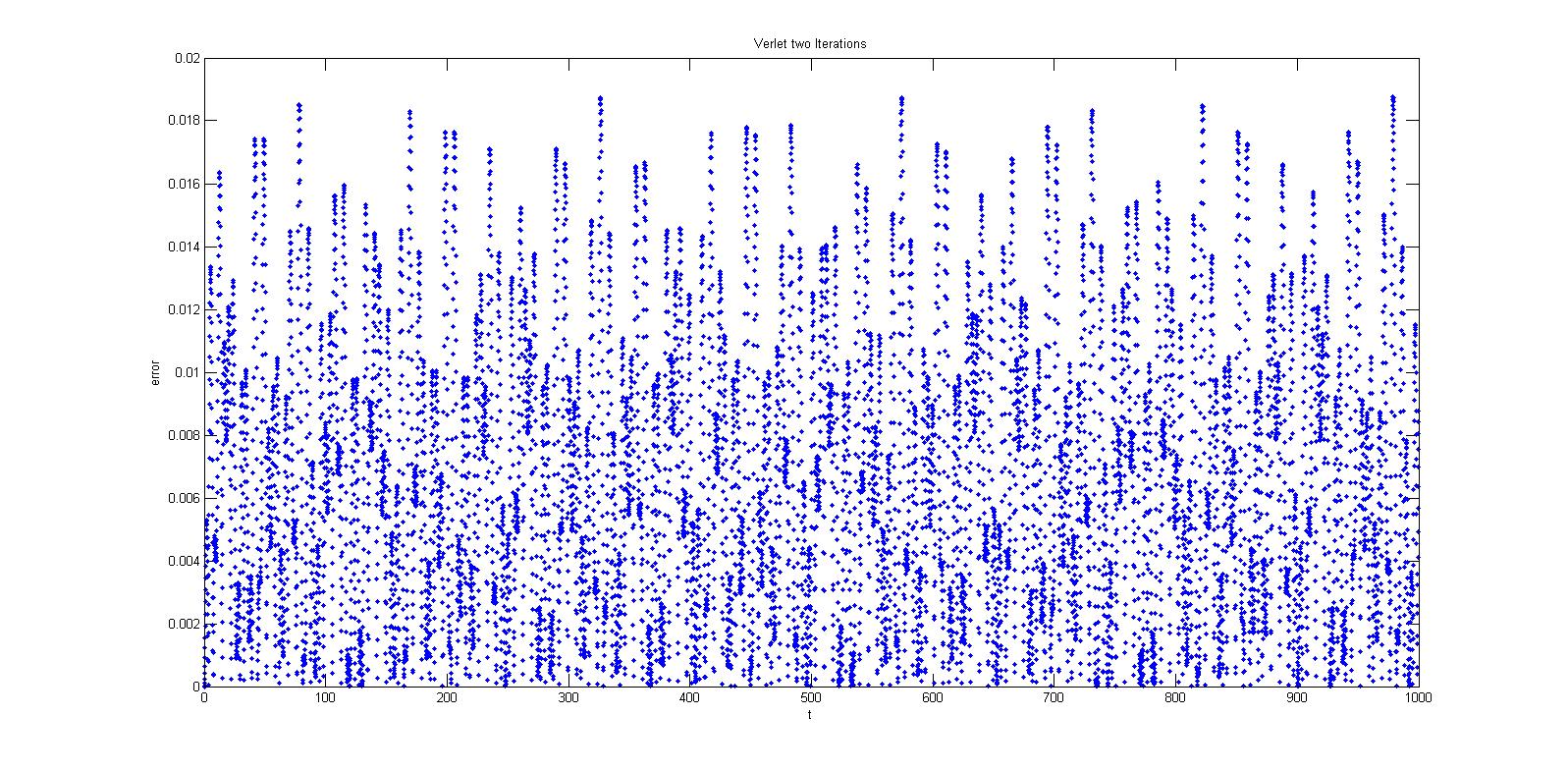}
	\end{minipage}
	
	\centering
	\includegraphics[width=0.45\textwidth]{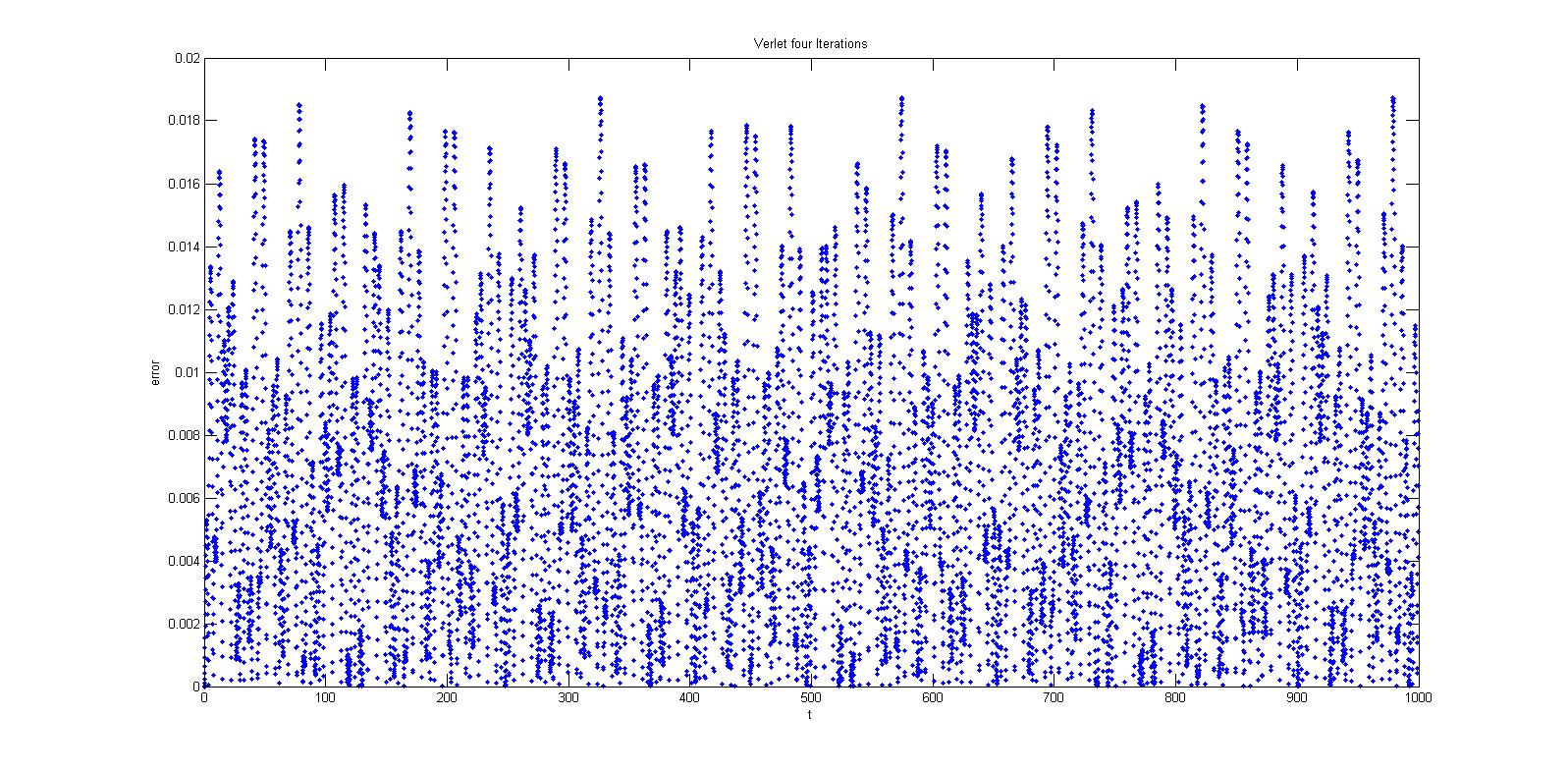}
	\caption{Errors of the numerical scheme: Iterative Verlet method.} 
 \label{fig:verlet_error}
\end{figure}

This were done for one, two and four iterations per timestep, to see whether how many iterations are reasonable. The results are shown in 
Figure \ref{fig:verlet_error}.

In a first comparison, we deal with the second order Verlet algorithm and
improve the scheme with iterative steps.

In a first initialisation process, one can see that the errors are very
similar to $1$ or $2$ iterative steps.

The reduction of the error is possible with the 
improvement to higher order initialisation scheme, e.g. start with a
first approximate solution with a RK scheme, or
apply extrapolation schemes. 

The following tables \ref{table1} and \ref{table2} should give an impression of the timescales of the problem and the errors. 
\begin{table}[ht]
\centering
\begin{tabular}{|c|c|c|c|}
\hline  \multicolumn{4}{|c|}{Runge-Kutta}  \\
\hline timestep & $10^{-5}$ & $10^{-3}$ & $10^{-1}$ \\ 
\hline number of steps & 1000000000 & 10000000 & 100000 \\ 
\hline computing time & 119 min & 23 sec & 2 sec \\ 
\hline stability & ok & ok & ok \\
\hline
\end{tabular} 
\caption{Stability and Computational Time with 4th order explicit Runge-Kutta.}
\label{table1}
\end{table}

\begin{table}[ht]
\centering
\begin{tabular}{|c|c|c|c|}
\hline  \multicolumn{4}{|c|}{Verlet}  \\
\hline timestep & $10^{-6}$ & $10^{-6}$ & $10^{-6}$ \\ 
\hline iterations per step & 1 & 2 & 4 \\
\hline stability & ok & ok & ok \\
\hline computing time & 67min & 120min & 219min \\
\hline mean error & 0.068 & 0.068 & 0.068 \\
\hline maximal error & 0.0187 & 0.0188 & 0.0187 \\
\hline
\end{tabular} 
\caption{Stability and Computational Time with 2nd order Verlet Scheme.}
\label{table2}
\end{table}

\begin{remark}
Obviously the iterations does not improve the algorithm, when only using
a lower order initialisation. 
By the way, it is sufficient to apply one iterative step
in in comparison with the Runge-Kutta algorithm.

Also we have a benefit in reducing the computational time instead of 
applying only Runge-Kutta schemes.

\end{remark}

\begin{figure}[htbp]
\centering	
	\begin{minipage}[b]{0.47\textwidth}
	\centering 
	\includegraphics[width=\textwidth]{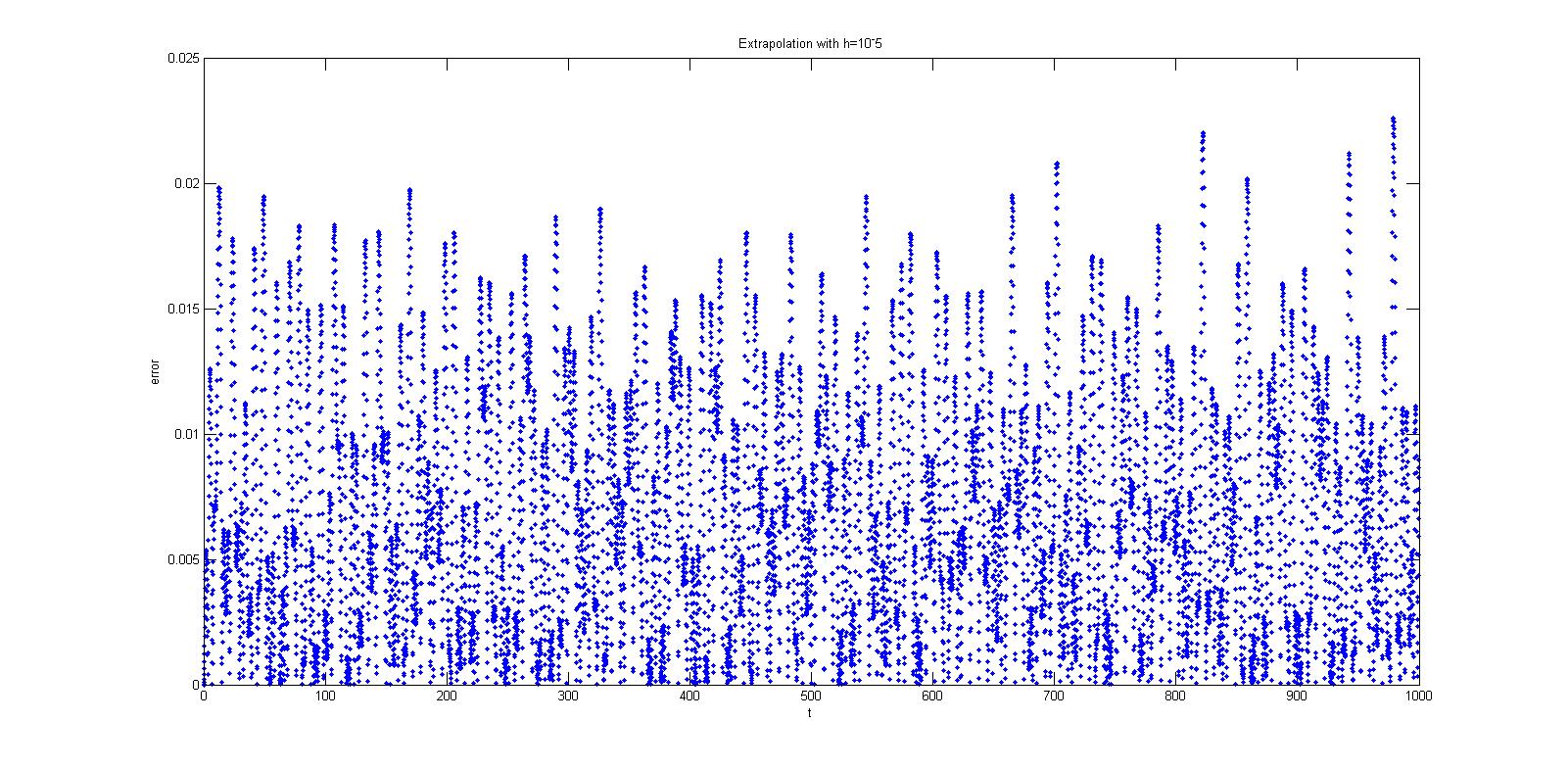}
	\end{minipage}
	\caption{Errors of the numerical scheme: 4th oder Extrapolation Scheme with Verlet method a Kernel ($h=10^{-5}$).} 
 \label{fig:extra_1}
\end{figure}

We tried to improve the solution with a extrapolation scheme in fourth order. 
We have a view at the errors this algorithm produces in comparison with the Runge-Kutta Solution with small time-steps ($10^{-5}$ time units per step).
In Figure \ref{fig:extra_1}, we presented the results of the 4th MPE 
method with different time-steps and compared it with the Runge-Kutta solution.

\begin{figure}[htbp]
\centering	
	\begin{minipage}[b]{0.47\textwidth}
	\centering 
	\includegraphics[width=\textwidth]{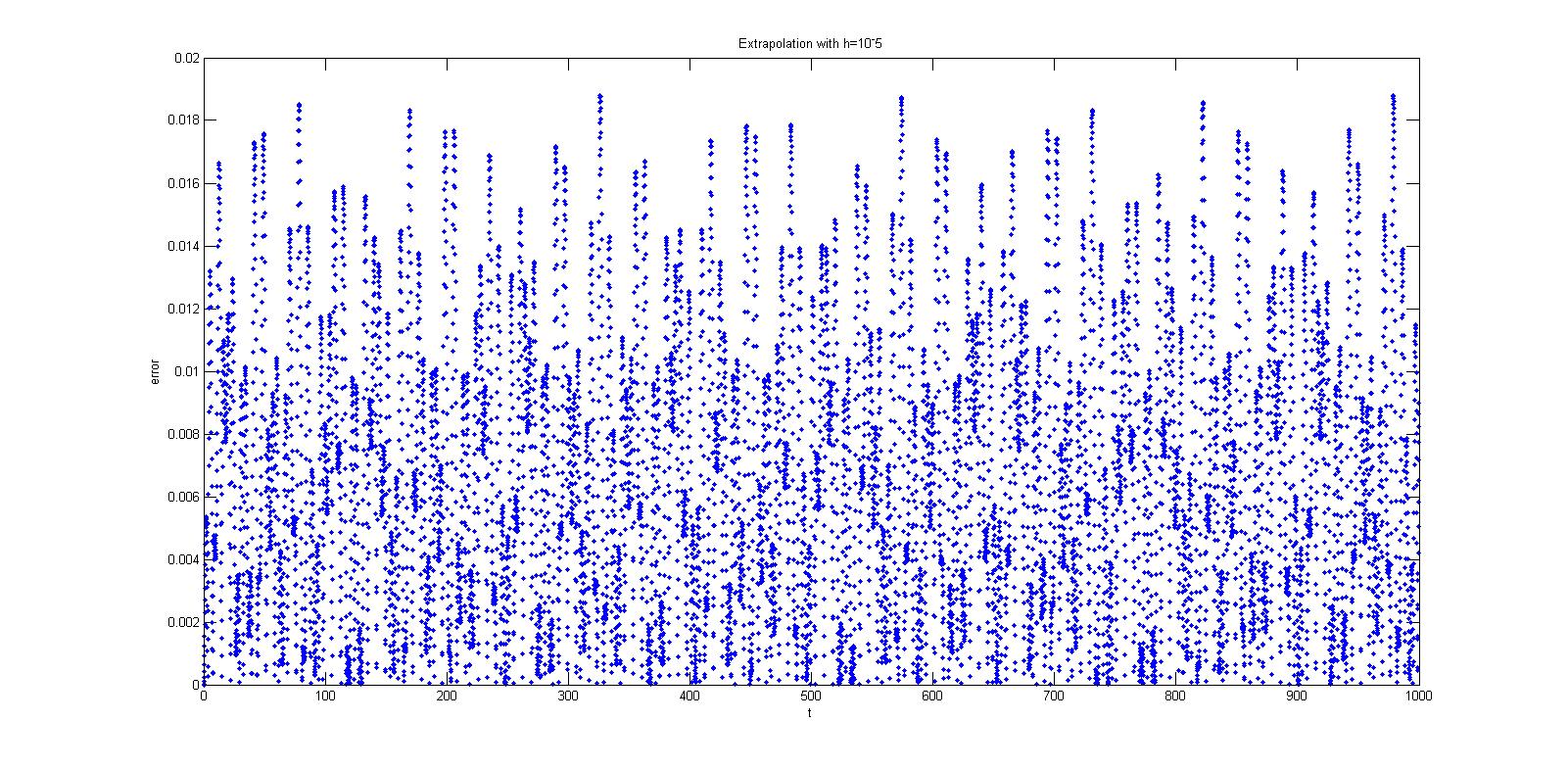}
	\end{minipage}
	\caption{Errors of the numerical scheme: 6th order Extrapolation Scheme with Verlet method a Kernel ($h=10^{-5}$).} 
 \label{fig:extra_2}
\end{figure}

Also we tested the 6th order MPE method with different time-steps and compared it with the Runge-Kutta solution, see Figure \ref{fig:extra_2}.

Like for Runge-Kutta we want to give an impression of the time scales for this extrapolation schemes, see Table \ref{table3}.
\begin{table}[ht]
\centering
\begin{tabular}{|c|c|c|c|c|}
\hline  & \multicolumn{2}{|c|}{Extrapolation 4th order} & \multicolumn{2}{|c|}{Extrapolation 6th order}  \\
\hline timestep &  $10^{-5}$ & $10^{-6}$ & $10^{-5}$ & $10^{-6}$   \\ 
\hline number of steps & 100000000 & 1000000000 & 100000000 & 1000000000 \\ 
\hline computing time & 14min & 142min & 29min & 272min  \\
\hline mean error & 0.007  & 0.007 & 0.0068 & 0.0068\\
\hline maximal error & 0.0226 & 0.0234 & 0.0188 & 0.0188\\
\hline
\end{tabular} 
\caption{Errors and Computational Time with 4th order MPE scheme unsing
Verlet Scheme as Kernel.}
\label{table3}
\end{table}

\section{Conclusions and Discussions}
\label{conc}

In the paper, we have presented a model to simulate a Levitron.
Based on the given Hamiltonian system, which is nonlinear, we 
present novel and simpler schemes based on splitting ideas to 
solve the equation systems.
In future, we concentrate on the numerical analysis and 
embedding higher order splitting kernels to the extraopoation schemes.

\section{Appendix}

\label{appendix}

For example,
the evolution of any dynamical variable $u(\bq,\bp)$ (including $\bq$ and $\bp$ themselves)
is given by the
Poisson bracket,
\begin{equation}
\pa_tu(\bq,\bp)=
                 \Bigl(
		          {{\partial u}\over{\partial \bq}}\cdot
                  {{\partial H}\over{\partial \bp}}
				 -{{\partial u}\over{\partial \bp}}\cdot
                  {{\partial H}\over{\partial \bq}}
				                    \Bigr)=(A+B)u(\bq,\bp).
\label{peq}
\end{equation}
For a separable Hamiltonian, 
\begin{equation}
H(\bp,\bq)={\bp^2\over{2m}}+V(\bq),
\label{ham}
\end{equation}
$A$ and $B$ are Lie operators, or vector fields
\be
A=\bv\cdot\frac{\pa}{\pa\bq} \qquad B=\bac(\bq)\cdot\frac{\pa}{\pa\bv}
\la{shop} 
\ee
where we have abbreviated $ \frac{\partial H}{\partial \bp}(\bp,  \bq) = \bv=\bp/m$ and $ - \frac{\partial H}{\partial \bq}(\bp,  \bq) = \bac(\bq)=-\nabla V(\bq)/m$.
The exponential operators $\e^{h A}$ and $\e^{h B}$ are then just shift operators. \\

 $S(h) = \e^{h/2 B} \e^{h A}  \e^{h/2 B}$  \\

That is also given as a Verlet-algorithm in the following scheme.

We start with $(\bq_0, \bv_0)^t = (\bq(t^{n}), \bv(t^{n}))^t $:

\begin{eqnarray}
(\bq_1, \bv_1)^t = \e^{h/2 B} (\bq_0, \bv_0)^t & = & ( I + \frac{1}{2} h \sum_i a(\bq) \frac{\partial}{\partial \bv_i}) (\bq_0, \bv_0)^t \\
& = & (\bq_0, \bv_0 + \frac{1}{2} h a(\bq_0) )^t ,
\end{eqnarray}

\begin{eqnarray}
(\bq_2, \bv_2)^t = \e^{h A} (\bq_1, \bv_1)^t & = & ( I + h \sum_i \bv_i \frac{\partial}{\partial \bq_i}) (\bq_1, \bv_1)^t \\
& = & (\bq_1 + h \bv_1 , \bv_1 )^t , 
\end{eqnarray}

\begin{eqnarray}
(\bq_3, \bv_3)^t = \e^{h/2 B} (\bq_2, \bv_2)^t & = & ( I + \frac{1}{2} h \sum_i a(\bq) \frac{\partial}{\partial \bv_i}) (\bq_2, \bv_2)^t \\
& = & (\bq_2, \bv_2 + \frac{1}{2} h a(\bq_1) )^t .
\end{eqnarray}

And the substitution is given the algorithm for one time-step $n \rightarrow n+1$:
\begin{eqnarray}
(\bq_3, \bv_3)^t = ( \bq_0 + h \bv_0 + \frac{h^2}{2} a(\bq_0) ,  \bv_0 + \frac{h}{2} a(\bq_0) + \frac{h}{2} a(\bq_0 + h \bv_0 + \frac{h}{2} a(\bq_0)) )^t ,
\end{eqnarray}
while $(\bq(t^{n+1}), \bv(t^{n+1}))^t = (\bq_3, \bv_3)^t$. \\

{\bf Iterative Verlet Algorithm}

In the abstract version of  
$\frac{\partial H}{\partial \bp}(\bp_{i-1}, \bq_{i-1}q)$, $- \frac{\partial H}{\partial \bq}(\bp_{i-1},  \bq_{i-1})$.

\begin{algorithm}

We have the iterative Verlet Algorithm:

1.) We start with the initialisation :  $(\bp_{0}(t^{n+1}), \bq_{0}(t^{n+1}))^t = (\bp(t^{n}), \bq(t^{n}))^t $ and $i=0$

2.) The iterative step is given as: $i=i+1$ and we have:
\begin{eqnarray}
\bq_{i}^{n+1} = \bq^n + h \frac{\partial H}{\partial \bp}(\bp^n - \frac{1}{2} h  \frac{\partial H}{\partial \bq}(\bp_{i-1}^{n+1}, \bq_{i-1}^{n+1}), \bq_{i-1}^{n+1}) ,
\end{eqnarray}
\begin{eqnarray}
&& \bp_{i}(t^{n+1}) =  \bp^n - \frac{h}{2} \frac{\partial H}{\partial \bq}(\bp_{i-1}^{n+1}, \bq_{i}^{n+1}) , \\
&& \bp_{i}(t^{n+1}) =  \bp^n - \frac{h}{2} \frac{\partial H}{\partial \bq}(\bp_{i-1}^{n+1},  \bq^n + h \frac{\partial H}{\partial \bp}(\bp^n - \frac{1}{2} h  \frac{\partial H}{\partial \bq}(\bp_{i-1}^{n+1}, \bq_{i-1}^{n+1}), \bq_{i-1}^{n+1}) ) , \nonumber 
\end{eqnarray}

where $h = t^{n+1} - t^n$ is the local time-step.

We compute the stopping criterion:

$\max(||\bp(t_{i}^{n+1} - \bp_{i-1}^{n+1} || , ||\bq(t_{i}^{n+1} - \bq_{i-1}^{n+1} ||) \le err$ or we stop after i= I, while $I$ is the maximal iterative step.

3.) The result is given as:

 $(\bp(t^{n+1}), \bq(t^{n+1}))^t = (\bp_i(t^{n+1}), \bq_i(t^{n+1}))^t $ 

and $n = n+1$  if $n > N$, while $N$ is the maximal time-step, we stop

else we go to step 1.)

\end{algorithm}

\bibliographystyle{plain}

\begin{thebibliography}{10}

\bibitem{chin2011}
S.~Chin and J.~Geiser. 
\newblock {\em Multi-product operator splitting as a general method of solving autonomous and non-autonomous equations.}
\newblock IMA J. Numer. Anal., first published online January 12, 2011.

\bibitem{dav78}
B.~Davis.
\newblock {\em Integral Transform and Their Applications.}
\newblock Applied Mathematical Sciences, 25, Springer Verlag, New York, Heidelberg, Berlin, 1978 .

\bibitem{dull98}
H.R.~Dullin and R.~Easton. 
\newblock{\em Stability of Levitron.}
\newblock Physica D: Nonlinear Phenomena, vol. 126, no. 1-2,  1-17, 1999.

\bibitem{EN00}
K.-J. Engel, R. Nagel,
\newblock {\em One-Parameter Semigroups for Linear Evolution Equations}.
\newblock Springer-Verlag, Heidelberg, New York, 2000.


\bibitem{fargei05}
I.~Farago and J.~Geiser.
\newblock{\em Iterative Operator-Splitting Methods for Linear Problems.}
\newblock  Preprint No. 1043 of the Weierstrass Institute for Applied Analysis and Stochastics, (2005) 1-18. International Journal of Computational Science and Engineering, accepted September 2007. 

\bibitem{gans97}
R.F.~Gans, T.B.~Jones, and M.~Washizu. 
\newblock{\em Dynamics of the Levitron.}
\newblock J. Phys. D., 31, 671-679, 1998.

\bibitem{gei-08}
J.~Geiser.
\newblock{\em Higher order splitting methods for differential equations: Theory and applications of a fourth order method.  Numerical Mathematics: Theory, Methods and Applications.}
\newblock Global Science Press, Hong Kong, China, accepted, April 2008. 

\bibitem{gei-08_2}
J.~Geiser and L.~Noack. 
\newblock{\em Iterative operator-splitting methods for nonlinear differential equations and applications of deposition processes}
\newblock Preprint 2008-4, Humboldt University of Berlin, Department of Mathematics, Germany, 2008.

\bibitem{goldstein81}
H.~Goldstein, Ch.P.~Poole, and J.~Safko.
\newblock{\em Classical mechanics}.
\newblock  Addison Wesley, San Francisco, USA, 2002.

\bibitem{hieb92}
M.~Hieber, A.~Holderrieth and F.~Neubrander.
\newblock{\em Regularized semigroups and systems of linear partial differential equations.}
\newblock Annali della Scuola Normale Superiore di Pisa - Classe di Scienze, Ser.4, 19:3, 363-379, 1992.


\bibitem{hoch05}
M.~Hochbruck and A.~Ostermann.
\newblock{\em Explicit Exponential Runge-Kutta Methods for Semilinear Parabolic Problems.}
\newblock SIAM Journal on Numerical Analysis, 43:3, 1069-1090, 2005.


\bibitem{han08}
E.~Hansen and A.~Ostermann.
\newblock {\em Exponential splitting for unbounded operators.}
\newblock Mathematics of Computation, accepted, 2008.


\bibitem{hil74}
Hildebrand, F.B. (1987) {\it Introduction to Numerical Analysis}. Second Edition, Dover Edition.

\bibitem{jan00} 
T.~Jahnke and C.~Lubich.
\newblock {\em Error bounds for exponential operator splittings.}
\newblock BIT Numerical Mathematics, 40:4, 735-745, 2000.

\bibitem{kan03} J.\,Kanney, C.\,Miller and C.\, Kelley.
\newblock {\em Convergence of iterative split-operator approaches for
approximating nonlinear reactive transport problems.}
\newblock Advances in Water Resources, 26:247--261, 2003.

\bibitem{kelly95}
C.T.~Kelly.
\newblock{\em Iterative Methods for Linear and Nonlinear Equations}.
\newblock Frontiers in Applied Mathematics, SIAM, Philadelphia, USA, 1995.


\bibitem{Mar68}
G.I ~Marchuk.
\newblock {\em Some applications of splitting-up methods
to the solution of problems in mathematical physics.}
\newblock Aplikace Matematiky, 1, 103-132, 1968.

\bibitem{stra68}
G.~Strang.
\newblock {\em On the construction and comparison
of difference schemes}.
\newblock SIAM J. Numer. Anal., 5, 506-517, 1968.

\bibitem{najfeld95}
I.~Najfeld and T.F.~Havel.
\newblock {\em Derivatives of the matrix exponential and their computation}.
\newblock Adv. Appl. Math, ftp://ftp.das.harvard.edu/pub/cheatham/tr-33-94.ps.gz, 1995. 


 \bibitem{strang68} 
Strang, G. (1968) On the construction and comparison of difference schemes. {\it SIAM J. Numer. Anal.}, {\bf 5}, 506-517.

\bibitem{suzu93}
M.~Suzuki.
\newblock{\em General Decomposition Theory of Ordered Exponentials.}
\newblock Proc. Japan Acad., 69, Ser. B, 161, 1993.


\bibitem{goldstein}
H.~Goldstein.
\newblock {\em Klassische Mechanik}.
\newblock Akademische Verlagsgesellschaft, Wiesbaden, 1981.


\end{thebibliography}

\end{document}